\documentstyle{article} 
\begin{document} 
\title{Embedding of Relativistic Particles \\ and Spinor Natural-Frame} 
\author{Naohisa OGAWA \thanks{E-mail: ogawa@particle.phys.hokudai.ac.jp} \ \ 
\\
Department of Mathematics, Hokkaido University, Sapporo 060 Japan}
\date{March 30, 1997} 
\maketitle 

\begin{abstract}
Embedding of Klein-Gordon and Dirac particle onto Riemannian submanifold in higher dimensional Minkowski space is given by using Hamiltonian BRST formalism.
 Up to the ordering and quantum potential term induced by embedding, obtained K-G equation is the usual one in Riemannian space, instead, the obtained Dirac equation is essentially different from the usual well known form using vierbein. The requirement of equivalence between two Dirac equations gives the property of natural-frame for spinor.
\end{abstract}

\section{Introduction}
  To consider the dynamics in gravity, usually we replace the dynamical equation generally covariant form.  Especially for the fermions, 
we put it at the Local Lorentz frame, that is, vierbein formalism.  
This is due to the essence of General Relativity: principle of general relativity and equivalence principle.  However,  we have another possibility to introduce the Riemannian manifold for dynamics. This is the case that our space-time is embedded in higher dimensional space-time in non trivial way.  In this case  space-time has larger dimension than four essentially, and our observation of space-time is only a part of it.
On the other hand,  The discussions of embedding into submanifold and its quantum effects are already given in details  for the non-relativistic particles with and without spin, \cite{constraint}, \cite{constraint2}, \cite{gauge} but there's nothing for relativistic ones.  In this paper, we treat the relativistic particles in larger dimensional space-time and taking the constraint for them to the Riemannian(Lorentzian) submanifold embedded in that space. (without touching its physical mechanism) The analysis is done by Hamiltonian BRST formalism \cite{FV}, \cite{battle}, \cite{ogawa} since there is time-reparametrization invariance, and for embedding constraint we utilize Faddeev-Senjanovic path-integral formulation equivalent to Dirac's procedure.  The discussion of quantum potential related to the extrinsic curvature \cite{constraint2} is not given here since we use the path-integral formulation, the reason will be given in detail at the time. And also we have no geometrical connection \cite{gauge} since our treatment is essentially due to the Dirac's procedure.  However, even for our case, the difference appears for Dirac equation between our embedding approach and usual vierbein approach \cite{Davis} in Riemannian manifold. The requirement of equivalence between these two equations gives the relation of these two wave functions which we call spinor-natural frame, and also proves the embedding hypothesis used in ref.\cite{ogawa.1}.

\section{Embedding of Spinless Relativistic Particle}
   Let us consider the classical canonical action for the relativistic free particle in 
D-dimensional Minkowski space time.
\begin{equation}
  S = \int d\tau~ [~p_a\dot{x}^a~ - ~N (p^2 - m^2)~],
\end{equation}
where, $\tau$ is a proper time, m is a mass, and N is a multiplier field.
From the Euler-Lagrange equations  we obtain the same equations as the ones obtained by the usual free relativistic particle's action as
\begin{equation}
  S' = - m \int d\tau ~ [ ~\eta_{ab}~ \dot{x}^a \dot{x}^b ~]^{1/2}.
\end{equation}
In this sense they are equivalent classically.
For simplicity we use the canonical action in the following.
To quantize this system, it is favorable to use the Hamiltonian BRST formalism \cite{FV} since it includes the local symmetry: time-reparametrization invariance.  This kind of work is already done by C.Battle, J.Gomis, and J.Roca \cite{battle}, but for our purpose we must extend it to confine the system into submanifold. For the concrete understanding we perform the calculation from the beginning, but in more intuitive way of Hamiltonian BRST formalism \cite{ogawa}. The action has the BRST invariance as
\begin{equation}
\delta^B x^a= 2 p^a c,~~\delta^B p^a = 0,~~\delta^B N = \dot{c},
\end{equation}
where $c$ is the Faddeev-popov ghost, and to insure the nilpotency we have the BRST transformation for ghost: $$ \delta^B c = 0.$$
Then the BRST-charge may take the form
\begin{equation}
Q_B = (p^2-m^2)c ~+~ \cdots,
\end{equation}
without the dotted terms we can not construct the BRST transformation for $N$. Note that 
$(p^2-m^2)$ is the first class constraint which induces the gauge transformation. To obtain the total BRST-charge, we must extend phase space to include the canonical pair of $N$ and $c$, that is, $\pi,~\bar{c}$, otherwise we can not construct $\delta^B N$.
But it is not enough from the following two reasons. 
\begin{enumerate}
\item Since the multiplier field $N$ become dynamical by introducing the canonical pair of $N$, we must kill its additional dynamical degree  by some ghost field, and construct the BRST-quartet mechanism.
\item It is impossible to introduce the time derivative of the field by canonical transformation. Therefore the BRST transformation of $N$ is not realized directly, but it is possible to realize on on-shell condition.
\end{enumerate}
From the above two reasons, we introduce another ghost pair $(\tilde{c},~~\bar{\tilde{c}})$, and require
\begin{equation}
 \{\tilde{c},~~\bar{\tilde{c}}\} = 1, 
\end{equation}
and the equation of motion $\tilde{c} = \dot{c}$. 
Then we take
\begin{equation}
Q_B = (p^2-m^2)c ~+~ \pi \tilde{c},
\end{equation}
which produce the BRST transformation for $N$ as
\begin{equation}
 \delta^B N = \{N, Q_B\} = \tilde{c} \approx \dot{c},
\end{equation}
where $\approx$ means equality holds only on equation of motion.
Since the phase space is extended, and new equation for ghost is required, it is necessary to add the following terms to our Lagrangian.
\begin{equation}
\Delta L = \pi \dot{N} + \bar{c}~(\dot{c}-\tilde{c}) + \bar{\tilde{c}}\dot{\tilde{c}}.
\end{equation}
This is all what we usually do in Hamiltonian-BRST formalism. The total Lagrangian takes the form
\begin{eqnarray}
L_{tot} &=& p_a\dot{x}^a+\pi \dot{N} + \bar{c}~(\dot{c}-\tilde{c}) + \bar{\tilde{c}}\dot{\tilde{c}} - N (p^2 - m^2) \nonumber \\
 &=& p_a \dot{x}^a+\pi \dot{N} + \bar{c}\dot{c} + \bar{\tilde{c}}\dot{\tilde{c}} - \{Q_B, N \bar{c}\}.
\end{eqnarray}
The BRST invariance of Lagrangian is trivial from the nilpotency and its canonical structure. The additional degree of freedom is found not to be physical since $\bar{\tilde{c}}, \pi, \tilde{c}, N$ are constructing the BRST-quartet. Then the Kernel is determined as
\begin{eqnarray}
&K(x_f \mid x_i)& = \int {\cal D}x ~{\cal D}p~ {\cal D}N ~{\cal D}\pi ~{\cal D}c ~{\cal D}\bar{c} ~{\cal D}\tilde{c} ~{\cal D}\bar{\tilde{c}} \nonumber \\
 && \exp \frac{i}{\hbar} \int^1_0 d\tau [p_a\dot{x}^a+\pi \dot{N} + \bar{c}\dot{c} + \bar{\tilde{c}}\dot{\tilde{c}} - N (p^2 - m^2) -\bar{c} \tilde{c} ~].
\end{eqnarray}
From the decoupling of ghost fields we can avoid them, and the Kernel takes the well known form as
\begin{equation}
K(x_f \mid x_i) = \int d^Dp \, \frac{1}{p^2-m^2} \, e^{\frac{i}{\hbar} p_a(x^a_f - x^a_i)}.
\end{equation}
Now we confine the above spinless particle onto the D-1 dimensional hyper surface embedded in our D dimensional Minkowski space.
The simplest way is to add the constraint $f(x) =0$ which specifies the hyper surface to the above obtained effective Lagrangian.
\begin{equation}
 L_{eff} = p_a\dot{x}^a+\pi \dot{N} - N (p^2 - m^2) + \lambda f(x),
\end{equation}
where we have neglected the ghost terms because they decouple with matter fields, and $\lambda$ is the multiplier. According to the Dirac's treatment of second class constraint, we obtain 4 second class constraints:
$$ \phi_1 = p_\lambda,~~\phi_2 = f(x),~~\phi_3 = p^a\partial_af,~~\phi_4 = \lambda(\partial f)^2 + 2N p^a p^b \partial_a \partial_b f,$$
and we obtain $$det^{1/2}\{\phi_i, \phi_j \} = (\partial f)^4.$$
Then the Kernel of this system can be obtained by using Faddeev-Senjanovic path-integral form as
\begin{eqnarray}
&&K(x_f \mid x_i) = \int {\cal D}x ~{\cal D}p~ {\cal D}N ~{\cal D}\pi ~{\cal D}\lambda ~{\cal D}p_\lambda ~(\partial f)^4 ~  \delta ( p_\lambda)~\delta ( f)~
\delta ( p^a\partial_af) \nonumber \\
 && \delta (\lambda(\partial f)^2 + 2N p^ap^b \partial_a\partial_b f) ~
e^{\frac{i}{\hbar} \int^1_0 d\tau [p_a\dot{x}^a+\pi \dot{N} - N (p^2 - m^2) + \lambda f(x) ]}.
\end{eqnarray}
The integration for $\lambda$ is performed easily, and we replace the coordinate to the one on hyper-surface and its normal one. We take the coordinate on hyper-surface as
$$ q^{\mu}~ =~ \{~ q^1,~ q^2, ~\cdots, ~q^{D-1} ~\},~~~
 with~its~conjugate~momentum~:~p_\mu,$$
and its normal coordinate as
$$ q^{\perp} \equiv f(x),~~~with~its~conjugate~momentum~:~p_\perp. $$
Since the coordinate transformation
\begin{equation}
 \{x^a, p_b\} \longrightarrow \{ (q^\mu, q^\perp), (p_\mu, p_\perp) \},
\end{equation}
is the point canonical transformation, the path-integral measure is preserved by Liouville theorem. The metric for the curvilinear coordinate is specified by
$$ \tilde{g}_{\mu \nu},~~ \tilde{g}_{\perp \perp}, ~~ \tilde{g}_{\mu \perp} = \tilde{g}_{\perp \mu} = 0,$$
and the metric on hyper surface is given by
$$ g_{\mu\nu} = \tilde{g}_{\mu\nu}\mid_{q^\perp=0} = \eta_{ab} \frac{\partial x^a}{\partial q^\mu} \frac{\partial x^b}{\partial q^\nu} \mid_{q^\perp=0},~~ g_{\perp \perp} =  \tilde{g}_{\perp \perp}\mid_{q^\perp=0} = \eta_{ab} \frac{\partial x^a}{\partial q^\perp} \frac{\partial x^b}{\partial q^\perp} \mid_{q^\perp=0}.$$
Then from 
$$ p_a =  \frac{\partial q^\mu}{\partial x^a} p_\mu + \frac{\partial q^\perp}{\partial x^a} p_\perp,$$
we obtain
$$p^a \partial_a f(x) = p^a \frac{\partial q^\perp}{\partial x^a} = p_\perp \tilde{g}^{\perp \perp}.$$
From above considerations, we can rewrite the Kernel into the form
\begin{eqnarray}
&&K(q_f \mid q_i) = \int {\cal D}q^\mu ~{\cal D}p_\mu~{\cal D}q^\perp ~{\cal D}p_\perp~ {\cal D}N ~{\cal D}\pi ~(\frac{\partial q^{\perp}}{\partial x^a})^2 ~\delta ( q^\perp)~
\delta ( p_\perp \tilde{g}^{\perp \perp}) \nonumber \\
 && \exp \frac{i}{\hbar} \int^1_0 d\tau [p_\mu \dot{q}^\mu + p_\perp \dot{q}^\perp + \pi \dot{N} - N ( \tilde{g}^{\mu\nu} p_\mu p_\nu  + \tilde{g}^{\perp \perp} p_\perp^2 - m^2) ].
\end{eqnarray}
Then from the $\pi$ integration, we find that N-integration is the usual (not path-) integration. Therefore we obtain
\begin{equation}
K(q_f \mid q_i) = \int {\cal D}q  \int^\infty_0  dN  \prod_\tau [ N^{-\frac{D-1}{2}} g^{1/2} ]~~ \exp \frac{i}{\hbar} \int^1_0 d\tau~ [ N m^2 + \frac{1}{4N} g_{\mu\nu} \dot{q}^\mu \dot{q}^\nu ]. 
\end{equation}
Next we change the time variable $\tau \to t = 2N\tau$,  with changing the infinitesimal time unit $\epsilon \to \varepsilon = 2N \epsilon$ in the path-integral measure in which $N^{-\frac{D-1}{2}}$ term is absorbed. By rewriting $T=2N$ we come to the final style:
\begin{eqnarray}
K(q_f \mid q_i) &=& [g(q_f) g(q_i)]^{1/4} \int^\infty_0  dT ~\int {\cal D}q ~  g^{1/2}(q)   \nonumber \\
&&  \exp \frac{i}{\hbar} \int^T_0 dt~ [~ \frac{1}{2} g_{\mu\nu} \dot{q}^\mu \dot{q}^\nu  + \frac{1}{2} m^2~]. 
\end{eqnarray}
we should give two comments at this stage.
Firstly, there is no $\hbar^2$ term in Lagrangian which is related to the ordering problem.
This is because we did not start from the operator formalism, and so its ordering is naturally fixed as Weyl-ordering when we use mid-point prescription. Second we do not obtain the  $\hbar^2$ quantum potential term related to the embedding.  The reason is the following.
By defining the natural frame as 
\begin{equation}
 f^a_\mu \equiv \frac{\partial x^a}{\partial q^\mu}\mid_{q^\perp=0},~~
 f^\mu_a \equiv  g^{\mu\nu} \eta_{ab} f^b_\nu,
\end{equation}
the quantum potential term usually obtained from the kinetic term in operator formalism as \cite{constraint2}
\begin{eqnarray}
 \eta^{ab} \{f^\mu_a,  p_\mu \} \{f^\nu_b,  p_\nu\} &=& (Laplace-Beltrami~operator) \nonumber \\
                              &&  + ~~(quantum~potential),
\end{eqnarray}
where $\{A,B \} \equiv (AB+BA)/2$, with taking care of the operator ordering.
But in our calculation we rewrite them just as $g^{\mu\nu}p_\mu p_\nu$ in the c-number and later we follow the Weyl-ordering. So we can not discuss quantum potential term in this calculation. The obtained formula for the Kernel is the same as usual propagator for spin-less relativistic particle in Riemannian manifold. Its derivation is written in the reference \cite{Schulman}. The simple introduction is the following. $m^2$ term in Klein-Gordon equation is replaced by auxiliary time derivative, and obtained equation has the similar form as non-relativistic Schr\"odinger equation of which Kernel is easily obtained.  The real propagator is obtained by Fourier-transformation by auxiliary time from that one. This ``time'' integration is the T-integration in our form.

\section{Embedding of Dirac particle}
Next we consider the embedding of Dirac particle as we have done in the previous section.
The classical action for Dirac particle is usually given in the form \cite{dirac},\cite{battle}
\begin{equation}
S_0 = \frac{1}{2}\int^1_0 d\tau [~ \frac{\dot{x}^2}{e} + em^2 + \chi \psi_a \dot{x}^a /e -\psi_a \dot{\psi}^a + \psi_S \dot{\psi_S} - m \psi_S \chi~],
\end{equation}
where $\psi_a,~ \psi_S,~ \chi$ are Grassmann odd, and other variables are even.
This system has two local symmetries. One is the time-reparametrization:
\begin{eqnarray}
&& \delta x^a = \xi \dot{x}^a,~ \delta e = \frac{d}{d\tau}(e \xi), \nonumber \\
& & \delta \psi_a = \xi \dot{\psi}_a,~ \delta \psi_S = \xi \dot{\psi}_S,~  \delta \chi = \frac{d}{d\tau}(\chi \xi),
\end{eqnarray}
where $\xi$ is the Grassmann even local parameter. Another one is the Super symmetry:
\begin{eqnarray}
&& \delta x^a = \alpha \psi^a, ~\delta e = -\alpha \chi, \nonumber \\
& & \delta \psi^a = \alpha (\dot{x}^a /e +\chi \psi^a/(2e)),~ \delta \psi_S = -\alpha, ~ \delta \chi = -2 \dot{\alpha},
\end{eqnarray}
where $\alpha$ is the Grassmann odd local parameter.
The related canonical action has the form:
\begin{equation}
S_c = \int^1_0 d\tau [~ p_a \dot{x}^a + \pi_a \dot{\psi}^a  + 
 \pi_S \dot{\psi}_S  - T_i N^i~],
\end{equation}
where the multiplier field:
\begin{equation}
  N^1 = e, ~~~~N^2 = \chi,
\end{equation}
and the 1st class constraints:
\begin{equation}
 T_1 = \frac{1}{2} (p^2 - m^2), ~~~~T_2 = \frac{1}{2}(\psi^a p_a + m \psi_S),
\end{equation}
The index 1 variables are Grassmann even (time-reparametrization), and index 2 variables are Grassmann odd (super transformation). 
Let us  introduce the ghost $c^i$ with opposite Grassmannian parity to $T_i$. The BRST transformation for each variables except $N^i$ are defined by charge $Q_B = T_i c^i$ as canonical transformation. Then $\delta_B N^i = \dot{c}^i$ is required from the invariance of action. To realize the BRST transformation for $N^i$ as before, we should introduce the one more pair of ghost field as,
\begin{equation}
   (\tilde{c}^{\,i},~\bar{\tilde{c}}_{\,j}), ~~~with~\{\tilde{c}^{\,i},~\bar{\tilde{c}}_{\,j}\}=\delta^i_j,
\end{equation}
and introduce the equation of motion as
\begin{equation}
 \tilde{c}^{\,i} = \dot{c}^i.
\end{equation}
Then the original BRST transformation is realized on on-shell condition, 
when we take $\delta_B N^i = \tilde{c}^{\,i}$.
To insure this transformation, we take
\begin{equation}
 Q_B = T_i c^i + \pi_i \tilde{c}^{\,i}.
\end{equation}
Then the additional term for Lagrangian is 
\begin{equation}
\Delta L = \pi_i \dot{N}^i + \bar{c}_i (\dot{c}^i - ~\tilde{c}^{\,i}) + \bar{\tilde{c}}_{\,i} \dot{\tilde{c}}^{\,i}.
\end{equation}
The total Lagrangian takes the form
\begin{eqnarray}
L &=&  p_a \dot{x}^a + \pi_a \dot{\psi}^a  + \pi_S \dot{\psi}_S +\pi_i \dot{N}^i + \bar{\tilde{c}}_{\,i} \dot{\tilde{c}}^{\,i} + \bar{c}_{i} \dot{c}^{i} - \bar{c}_{i} \tilde{c}^{\,i}- T_i N^i, \nonumber \\
&=& (Kinetic~terms) - \{ Q_B, N^i \bar{c}_i \},
\end{eqnarray}
 so is the BRST invariant from the nilpotency, and the gauge fixing is done automatically.
The ghost fields are decoupled, and so we can neglect them, and after the some trivial integrations, we come to the formula for Kernel
\begin{equation}
K(x_f \mid x_i) = \int {\cal D}x {\cal D}p \int d\theta \int^\infty_0 dT \exp 
\frac{i}{\hbar} \int^1_0 d\tau [~p_a \dot{x}^a - T_i N^i~],
\end{equation}
while deriving the above formula we have fixed $\psi_a$, and $\psi_S$ at $t=0,\, 1$ as the boundary condition, and $N^1 = T, ~ N^2 = \theta$. The remained integration is performed easily with Grassmannian integration for $\theta$ in the form:
\begin{equation}
K(x_f \mid x_i) = \int d^Dp  ~\frac{\psi_a p^a + m \psi_S}{p^2 - m^2} 
~ e^{\frac{i}{\hbar} p_a (x^a_f - x^a_i)},
\end{equation}
which is the well known form for Dirac propagator when we replace $ \psi^a \to \Gamma^a$ :gamma matrix, $\psi_S \to 1$. \cite{dirac}
Now let us consider the dimensional reduction: embedding as before.
Our effective Lagrangian for our purpose is
\begin{equation}
L_{eff} =  p_a \dot{x}^a - \frac{T}{2} (p^2-m^2) + \frac{\theta}{2} (\psi_a p^a + m \psi_S) + \lambda f(x).
\end{equation}
This second-class system induces the following 4-constraints,
$$ \phi_1 = p_\lambda,~~\phi_2 = f(x),~~\phi_3 = K^a\partial_af,~~\phi_4 = \lambda T  (\partial f)^2 + K^a K^b \partial_a \partial_b f,$$
where $$ K^a = T p^a - \frac{1}{2} \theta \psi^a, $$ 
and we obtain $$det^{1/2}\{\phi_i, \phi_j \} = T^2 (\partial f)^4.$$
The Faddeev-Senjanovic path-integral formula takes the form after some trivial integration
\begin{eqnarray}
K(x_f \mid x_i) &=& \int {\cal D}x~ {\cal D}p \int d\theta \int^\infty_0\!\! dT~ \prod_{\tau} \, [T(\partial f)^2] ~\delta{(f)}~\delta{(K^a \partial_a f)} \nonumber \\
&&\!\!\!\!\! \exp \frac{i}{\hbar} \int^1_0 d\tau \,[\,p_a \dot{x}^a - \frac{T}{2}(p^2-m^2) + \frac{\theta}{2}(\psi^a p_a + m\psi_S)\,].
\end{eqnarray}
Now we can rewrite this Kernel by using independent variables in quite the same way as we have done for spin-less particle.  The result is the following.
\begin{equation}
K(q_f \mid q_i)  =  \int {\cal D}q~ {\cal D}p \int d\theta \int^\infty_0 \!\! dT~e^{\frac{i}{\hbar} \int^1_0 d\tau \,[\,p_\mu \dot{q}^\mu - \frac{T}{2}(g^{\mu\nu} p_\mu p_\nu -m^2) + \frac{\theta}{2}(\Gamma^\mu(q) p_\mu + m)\,]}
\end{equation}
where $\Gamma_\mu = f^a_\mu(q) \Gamma_a$, and we have replaced $\psi_a \to \Gamma_a, ~~\psi_S \to 1$ at the last stage of calculation as we have done in Minkowski case. We may rewrite the above formula into the following form:
\begin{equation}
K(q_f \mid q_i)  =  \int {\cal D}q~ {\cal D}p \int^\infty_0 \!\! dT~e^{\frac{i}{\hbar} \int^1_0 d\tau \,[\,p_\mu \dot{q}^\mu - T(\Gamma^\mu(q) p_\mu - m) \,]}.
\end{equation}
We use the equivalence of these two types of Kernel without proof from the following reasons. First, the equivalence of these two forms is hold at least in Minkowski case. 
Second, they both have general covariance and hidden global Lorentz invariance, 
and include elements of same Clifford algebra. The Kernel here is easily found to have the same physics as the following type of Dirac equation as we have done for spin-less particle.
\begin{equation}
[\, i\Gamma^a f^\mu_a (q) \partial_\mu - m \,] \tilde{\Psi}(q) = 0.
\end{equation}
We call this Dirac equation as Embedding-Dirac equation. We should notice the following points. The equation has general coordinate transformation invariance, and the spinor is in the $SO(D-1,1)$ group defined in external space-time. In the same way the index ``a" is not the local Lorentz index but the global Lorentz one, therefore there is no spin-connection term. When we  consider the fermion in curved space-time, we usually put it on Local-Lorentz frame. But in our case, we have another possibility to treat fermion by assigning it the external dimensional spinor and constraining on the hypersurface as our curved manifold.  This is, however, not enough. The Gauss equation is given by
\begin{equation}
 \partial_\lambda f^\mu_a = -\Gamma^\mu_{\lambda \nu} f^\nu_a + H_\lambda^{~\mu} N_a,
\end{equation}
where $H_{\mu\nu}$ is the extrinsic curvature, and $N_a$ is the space like vector normal to our hypersurface. This equation helps us to calculate the divergence of current as
\begin{equation}
 \nabla_\mu [\, \bar{\tilde{\Psi}} \Gamma^\mu \tilde{\Psi}\,] ~=~ H^\mu_{~\mu} N_a \bar{\tilde{\Psi}} \Gamma^a \tilde{\Psi}.
\end{equation}
Thus the current conservation is violated by the extrinsic mean curvature term in tree level.
To obtain the current conservation in our manifold, we need the subsidiary condition.
\begin{equation}
   \bar{\tilde{\Psi}} \Gamma_\bot \tilde{\Psi}  = 0, ~~~~ 
\Gamma_\bot \equiv N_a \Gamma^a.
\end{equation}
The usual Dirac equation has the form
\begin{equation}
[\, i\gamma^\mu (\partial_\mu  + \frac{1}{2} \omega_\mu) - m \,] \Psi(q) = 0,
\end{equation}
while $\gamma^\mu = \gamma^i e^\mu_i$, where $\gamma^i$ is the $SO(D-2,1)$ 
$\gamma$-matrix defined on Local-Lorentz frame, $e^\mu_i$ the vierbein, 
and $\omega_\mu$ is spin-connection. We call this equation as LL-Dirac equation.
These two (large and small) gamma matrices satisfy the same algebra though they are originally in the different dimensional Lorentz group.
\begin{equation}
  \{\Gamma^\mu,~\Gamma^\nu \} = \{\gamma^\mu,~\gamma^\nu\} = 2g^{\mu\nu}.
\end{equation}
This means the natural frame gives the dimensional reduction of spinor algebra, and two kinds of spinors have the same degree of freedom though they are in the different dimensional one. Therefore above two Dirac equations should be physically equivalent, and only the representation is different. If we require this equivalence, we obtain the relation between these two kinds of spinor, and we are led to the embedding hypothesis as the natural consequence. \cite{ogawa.1}

\section{Spinor Natural-Frame}
The usual natural-frame is defined between vector on external coordinate and internal one. Similarly we can define the natural-frame between external and internal spinor, which we call Spinor natural-frame: ${\cal F}$ as
\begin{equation}
   \Psi(q) ~= ~{\cal F}(q)~ \tilde{\Psi}(q).
\end{equation}
Firstly we assume the existence of the left inverse of ${\cal F}$ in the form
\begin{equation}
         {\cal G} = A {\cal F}^{\dag}, ~~~~  {\cal G}{\cal F} = 1,
\end{equation}
where $A$ is some unknown matrix. This is in other words the assumption for the existence of left inverse for ${\cal F}^{\dag} {\cal F}$. Spinor natural-frame should satisfy some conditions which are required from the equivalence of two Dirac equations. By putting (43) into LL-Dirac equation, multiplying ${\cal F}^{\dag}$ from the left, and we obtain the necessary condition to obtain the Embedding-Dirac equation as,
\begin{eqnarray}
&&{\cal F}^{\dag}{\not D} {\cal F} \equiv {\cal F}^{\dag} \gamma^\mu D_\mu {\cal F} \equiv {\cal F}^{\dag} \gamma^\mu (\partial_\mu + \frac{1}{2} \omega_\mu) {\cal F} = 0, \\
&& {\cal F}^{\dag} \gamma^\mu {\cal F} = {\cal F}^{\dag} {\cal F} \, \Gamma^\mu.
\end{eqnarray}
Further $\bar{\tilde{\Psi}}\tilde{\Psi}$ and $\bar{\Psi}\Psi$ are both Lorentz and diffeomorphism scalar, and both are related to the particle's probability density.
Therefore it is natural to require the condition,
\begin{equation}
\bar{\tilde{\Psi}}\tilde{\Psi} = \bar{\Psi}\Psi.
\end{equation}
We should notice that this condition is necessary to obtain the equivalence of two actions related to two Dirac equations. Also it is consistent with the result of section 2, where the embedding treatment of scalar wave function and usual scalar wave function in Riemannian manifold satisfy the same equation of motion. By taking the same normalization, we have $\tilde{\Phi} = \Phi$  which is equivalent to the above requirement.
This condition is rewritten in the form:
\begin{equation}
 \Gamma^0 = {\cal F}^{\dag} \gamma^0 {\cal F},
\end{equation}
where each index ``0"  is the global and local Lorentz index respectively but not coordinate index. By multiplying $e^0_\mu$ in the second condition, and the use of the third condition gives 
\begin{equation}
\Gamma^0 {\not \Gamma} = {\cal F}^{\dag}{\cal F}, ~~~ 
1 = {\not \Gamma} \, \Gamma^0 ({\cal F}^{\dag}{\cal F}), ~~~
 {\not \Gamma} \equiv \Gamma^a  f^\mu_{~a} e^0_\mu,
\end{equation}
where we used $(\Gamma^0)^2 = 1$ and ${\not \Gamma}^2 = \eta^{00} =1$.
 We obtain the explicit form of ${\cal G}$.
\begin{equation}
{\cal G} = {\not \Gamma} \, \Gamma^0 {\cal F}^{\dag},~~~ 
1 = ({\cal F}^{\dag}{\cal F}) {\not \Gamma} \, \Gamma^0,
\end{equation}
These equations show the existence of the inverse of ${\cal F}^{\dag} {\cal F}$. 
By multiplying ${\cal G}$ from left on (43), we obtain
\begin{equation}
  \tilde{\Psi}(q) = \,{\cal G} \Psi(q),
\end{equation}
which is the inverse relation of (43). 
The right inverse of ${\cal F}$ is found to be ${\cal G}$ when $({\cal F} \Gamma^0 {\cal F}^{\dag})^2 = 1$ holds. Then we take
\begin{equation}
\gamma^0 = {\cal F}\, \Gamma^0 {\cal F}^{\dag},
\end{equation}
which is the another requirement for ${\cal F}$. 
Now ${\cal F}$ and ${\cal G}$ are inverse each other, so we can rewrite the first two conditions into the form 
(by multiplying ${\cal F} \, {\not \Gamma} \, \Gamma^0$ from left),
\begin{equation}
{\not D} {\cal F} = 0, ~~~~ \gamma^\mu {\cal F} = {\cal F} \Gamma^\mu.
\end{equation}
Then by putting (51) into (37), and multiplying ${\cal F}$ from the left, this should reduce to the usual L.L.Dirac equation.
This requirement takes the following form.
\begin{equation}
  {\cal F}\, \Gamma^\mu {\cal G} = \gamma^\mu, ~~~ {\cal F}\, \Gamma^\mu \partial_\mu {\cal G} = \frac{1}{2} \gamma^\mu \omega_\mu.
\end{equation}
The first condition is automatically satisfied from (53), and the second one is reduced to 
$0= ({\not D} {\cal F}) \, {\cal G}$ by Leibnitz rule, and is satisfied by (53).
Lastly we consider the subsidiary condition. By using the above relations,
\begin{equation}
0 = \bar{\tilde{\Psi}} \Gamma_\bot \tilde{\Psi}  = \bar{\Psi} ({\cal F}\, \Gamma_\bot {\cal F}^{\dag}) \gamma^0 {\cal F}{\cal F}^{\dag} \gamma^0 \Psi.
\end{equation}
This relation requires the condition ${\cal F}\, \Gamma_\bot {\cal F}^{\dag} = 0.$ 
 Those obtained five conditions are necessary conditions to determine ${\cal F}$, which we write down here again.
\begin{eqnarray}
&&{\cal F}^{\dag}{\not D} {\cal F}=0,~~~{\cal F}^{\dag}\gamma^\mu {\cal F} = {\cal F}^{\dag}{\cal F} \Gamma^\mu,~~~
\Gamma^0 = {\cal F}^{\dag} \gamma^0 {\cal F}, \nonumber \\  
&& \gamma^0 = {\cal F}~ \Gamma^0 \,{\cal F}^{\dag},~~~{\cal F}\, \Gamma^a {\cal F}^{\dag} = f^a_\mu \gamma^\mu {\cal F}{\cal F}^{\dag},
\end{eqnarray}
where the last condition is the subsidiary one rewritten by using the relation 
$\Gamma^a = f^a_\mu \Gamma^\mu + N^a \Gamma_\bot$.
We should remark here that ${\cal F}{\cal F}^{\dag}$ is determined by last two equations in the form,
\begin{equation}
{\cal F}{\cal F}^{\dag} = {\not \gamma}^{\, (-1)} \gamma^0, ~~~ 
{\not \gamma} \equiv  f^0_\mu \gamma^\mu.
\end{equation}
  Next we consider the relation between vector fields for external and internal.
The vector field in Local-Lorentz frame can be defined as $\Phi^i \equiv \bar{\Psi} \gamma^i \Psi$. This vector field can also be constructed by using $\tilde{\Psi}$ as follows,
\begin{eqnarray}
\Phi^i &=& \bar{\Psi} \gamma^i \Psi = \tilde{\Psi}^{\dag} {\cal F}^{\dag} \gamma^0 \gamma^i {\cal F} \tilde{\Psi} = \tilde{\Psi}^{\dag} {\cal F}^{\dag} \gamma^0 {\cal F} \Gamma^{\mu} \tilde{\Psi} e^i_\mu , \nonumber \\
&=& \tilde{\Psi}^{\dag} \Gamma^0 \Gamma^{\mu} \tilde{\Psi} e^i_\mu = e^i_\mu f^\mu_a \tilde{\Phi}^a,
\end{eqnarray}
where $\tilde{\Phi}^a \equiv \bar{\tilde{\Psi}} \Gamma^a \tilde{\Psi}$ is the vector field with external space-time index. Inversely by using the subsidiary condition,
\begin{eqnarray}
\tilde{\Phi}^a &=& \bar{\tilde{\Psi}}\, \Gamma^a \tilde{\Psi}\, = \,
              f^a_\mu \, \bar{\tilde{\Psi}}\,  \Gamma^\mu \tilde{\Psi} \, = \, 
              f^a_\mu \bar{\Psi} \gamma^0 {\cal F} \Gamma^0 {\not \Gamma}^{\dag} \Gamma^0 \Gamma^\mu {\not \Gamma} \Gamma^0 {\cal F}^{\dag} \Psi \nonumber \\
 &=& f^a_\mu \, \bar{\Psi} \, \gamma^\mu \Psi \,= \, f^a_\mu \, e^\mu_i \, \Phi^i,
\end{eqnarray}
where we used the relation $\Gamma^0 {\not \Gamma}^{\dag} \Gamma^0 = {\not \Gamma}$. Thus we found the simple relation for outer and internal vector fields. 
We write down all together for scalar, spinor, vector transformation law.
\begin{equation}
\Phi = \tilde{\Phi},~~~\Psi ~= ~{\cal F}~ \tilde{\Psi},~(\tilde{\Psi} ~= ~{\cal G}{\Psi}),~~~\Phi^i = e^i_\mu f^\mu_a \tilde{\Phi}^a,~(\tilde{\Phi}^a = f_\mu^a e^\mu_i {\Phi}^i).
\end{equation}

\section{Discussion}
We have discussed the embedding of spin-less and spinning particle onto the sub-Riemannian manifold in relativistic formulation. The obtained equation on the Riemannian manifold is not the usual form in the case of spinning particle. By requiring the equivalence of these two equations, we obtain the relation between outer and internal spinor fields, and it shows also the relation between outer and internal vector fields.
Our discussion was based on the dimensional-difference =1, but it is straightforward to extend it to the general cases.  The quantum potential \cite{constraint},\cite{constraint2} 
 can also be calculated in this line by considering the relation with operator formalism, or by starting from the gauge fixed Lagrangian with constraint (12),(33), and quantizing in operator formalism, though we did not touch it in this paper.
The spinor natural-frame defined in this paper satisfies 5 conditions (56).
The existence of the solution for ${\cal F}$ is just assumed and is not discussed here.
 This is remained as open question. The discussion given in this paper is basing on the space-time embedding to hold the explicit covariance. But in the special case we can take the non-relativistic frame:
$$ f^a_\mu = (\, f^0_0 =1, f^0_{\mu=spacial} = f^{a=spacial}_0 =0 \,), $$
where we took the special Lorentz frame that our hypersurface is [external-time] $\otimes$  [D-2 dim. spacial-surface]. In such a frame, we can take also
$$ e^i_\mu = (\, e^0_0 =1, e^0_{\mu=spacial} = e^{i=spacial}_0 =0 \,), $$
and then $$\Phi^i = e^i_\mu f^\mu_a \tilde{\Phi}^aC~~~ \tilde{\Phi}^a = f_\mu^a e^\mu_i {\Phi}^i$$ holds for spacial index ``i" and ``a". This is the relation for non-relativistic vector fields, which is discussed 
in Ref.\cite{ogawa.1}.
\

\

\noindent{\em Acknowledgement.}\\
The author wishes to thank Prof.K.Fujii and Prof.T.Okazaki for their usual encouragement 
and discussions.

 \end{document}